\documentclass[12pt]{article}
\usepackage{array, booktabs}
\usepackage{bm}
\usepackage{graphicx}
\usepackage{caption}
\usepackage{amsmath,amssymb,wasysym}
\usepackage{algorithm}
\usepackage{multirow}
\usepackage{algpseudocode}
\usepackage{lineno}  
\usepackage{setspace}  
\usepackage{mathtools}

\usepackage[utf8]{inputenc}
\usepackage{xcolor}
\usepackage{hyperref}
\usepackage{cleveref}
\usepackage[margin=1 in]{geometry}

\usepackage[utf8]{inputenc}
\usepackage[T1]{fontenc}
\usepackage{helvet}

\usepackage[affil-it]{authblk}
\usepackage{lineno}  
\usepackage{setspace}  

\DeclareMathOperator\erfc{erfc}

\makeatletter
\newcommand*{\centerfloat}{%
  \parindent \z@
  \leftskip \z@ \@plus 1fil \@minus \textwidth
  \rightskip\leftskip
  \parfillskip \z@skip}
\makeatother

\usepackage{xcolor}

\usepackage[
    backend=biber,natbib,
    style=nature,
]{biblatex}
\begin{document}


\title{Nutrient Transport in  Concentration Gradients}
\author{Jingyi Liu$^1$, Yi Man$^2$, Eva Kanso$^{1,3}$\footnote{kanso@usc.edu} \\ 
\footnotesize{$^1$Department of Aerospace and Mechanical Engineering,  \\ University of Southern California, Los Angeles, California, USA} \\
\footnotesize{$^2$Mechanics and Engineering Science, Peking University, Beijing 100871, China} \\
\footnotesize{$^3$Department of Physics and Astronomy, University of Southern California, Los Angeles, California, USA}}


\date{\today}


\maketitle

Sessile ciliates attach to substrates and generate feeding currents to capture passing particulates and dissolved nutrients.  
Optimal ciliary activity that maximizes nutrient flux at the cell surface while minimizing the rate of hydrodynamic energy dissipation is well characterized in uniform nutrient fields. 
However, it is unclear how ciliary motion should change when nutrients are non-uniform or patchy. To address this question, we modeled the sessile ciliate and feeding currents using the spherical envelope model, and used an unsteady advection-diffusion equation to describe the nutrient scalar field.
In the absence of flows, we calculated the diffusive nutrient uptake analytically in linear nutrient gradients and found no advantage over uptake in uniform fields. With ciliary activity driving feeding currents, we used a spectral method to solve for the unsteady nutrient concentration. We found that, when the axis of symmetry of the ciliary motion is aligned with the nutrient gradient, 
nutrient uptake at the cell surface increases steadily over time, with highest uptake achieved by the treadmill ciliary motion which is optimal in uniform fields as well.
The associated nutrient uptake in concentration gradients scales with the square root of the product of time and Péclet number. In patchy environments,
optimal ciliary activity depends on the nature of the patchiness.
Our findings highlight strategies that enable sessile ciliates to thrive in  environments with fluctuating nutrient availability.

\spacing{1.5}


\section{Introduction}
Marine microbe often inhabit environments with patchy or non-uniform nutrient distribution~\cite{fenchel1999motile, long2001microscale, durham2013turbulence, keegstra2022ecological, stocker2012marine, babko2020oxygen, stal2022marine}. 
Motile microbes, from bacteria~\cite{stocker2012marine, clerc2022survival} to ciliates~\cite{raina2019role}, navigate toward nutrient-dense regions through chemotaxis, actively seeking out localized nutrient hotspots.
However, sessile ciliates attach to substrates~\cite{jonsson2004}
and employ coordinated ciliary movements to generate feeding currents and draw nutrients toward their feeding structures \cite{christensen2003}. The effectiveness of this sessile lifestyle in transporting nutrients in non-uniform and patchy environments is not well understood.

To survive in various environments, sessile ciliates have evolved diverse strategies; \textit{Vorticella}, for example, adjust their feeding currents by regulating their orientation relative to the substrate to which they attach~\cite{pepper2010, pepper2013, pepper2021effect}. They, and their \textit{Pseudovorticella} relatives also attach to suspended particles, forming moving clusters, presumably to enhance nutrient flux to the entire cluster and reduce their predation risk~\cite{krishnamurthy2023active,kanso2021teamwork}. \textit{Stentors} switch between sessile and free-swimming states to explore new environments~\cite{jonsson2004, slabodnick2017macronuclear,echigoya2022switching, shekhar2023cooperative}. 
These strategies reflect the adaptive mechanisms sessile ciliates use to generate flows. However, most studies focus on nutrient uptake in uniform nutrient fields~\cite{karp1996, solari2006multicellularity,short2006flows,michelin2011,michelin2013unsteady, liu2024feeding,liu2024optimal}, with little known about nutrient uptake in non-uniform nutrient fields. Do sessile ciliates improve or reduce their nutrient flux in concentration gradients compared to uniformly-rich nutrient backgrounds? We address this question using a mathematical model that couples cilia-driven flows with unsteady nutrient concentration.

We use Blake's envelope model~\cite{blake1971} to describe the fluid flow generated by the surface ciliary activity of a sessile ciliate. This model was used examine the effects of cilia-driven flows on nutrient transport in uniform nutrient fields \cite{magar2003, magar2005, michelin2013unsteady, liu2024feeding, liu2024optimal}. Here, we study the effects of these flows on the unsteady evolution of a non-uniform nutrient field. Specifically, we consider nutrient fields characterized by a linear concentration gradient (Fig.~\ref{fig: pure_diff}A).
 Our approach proceeds in three stages. First, in the absence of cilia-driven flows, we derive an analytical solution for the unsteady nutrient diffusion in a linear concentration gradient. Thus, we establish a baseline for understanding diffusive nutrient transport to the cell surface in a linear concentration field in the absence of flow. Next, we numerically solve, using a spectral method, the unsteady advection-diffusion equation to capture the effects of cilia-driven flows on nutrient uptake over time, identifying a scaling law that relates surface activity to optimal feeding in a linear concentration field. Finally, we extend our model to patchy concentration fields, exploring cilia-driven flows that significantly enhance nutrient intake. This analysis offers a quantitative framework for understanding how sessile ciliates optimize feeding in heterogeneous nutrient conditions and could have ecological implications on understanding diversity at the micron scale.


\section{Nutrient diffusion in linear concentration gradients}

\begin{figure*}[tbhp!]
		\centerline{\includegraphics{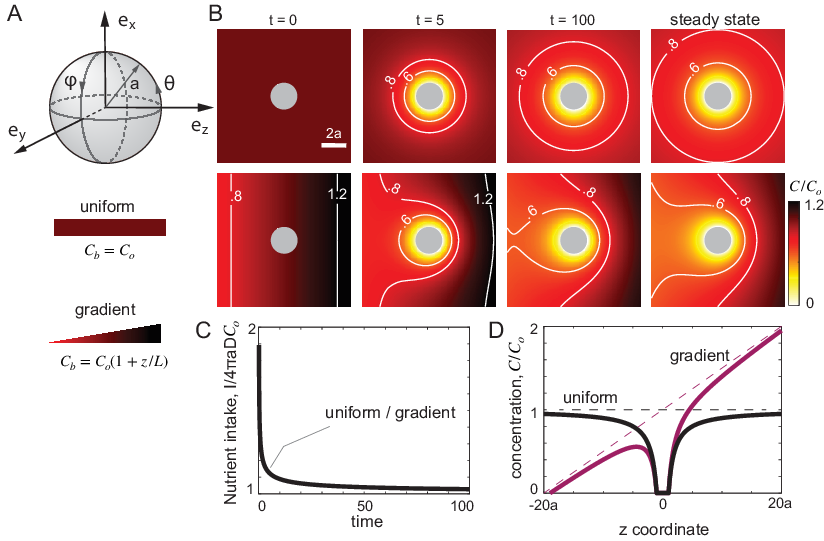}}
	\caption[]{\textbf{Pure diffusion in concentration gradient.} A. Mathematical model of sessile ciliated cell as a sphere of radius $a$, with spherical  $(r, \theta, \phi)$ and Cartesian $(x,y,z)$ coordinates. B. Concentration field of dissolved nutrients following a diffusion process in an initially (top row) uniform concentration ($\epsilon = 0$) and (bottom row) gradient ($\epsilon = 0.05$). Snapshots shown at time instances $t = [0, 5, 100]$ and at steady-state. C. Nutrient intake $J=I/4\pi aD C_o$ in pure diffusion in uniform and linear gradients. D. Steady-state solution of concentration $c(x=0,y=0,z)$ as a function of the $z$ coordinate. }
	\label{fig: pure_diff}
\end{figure*}

\paragraph{Problem setting} We describe the concentration $C(t, r,\theta, \phi)$ of nutrients around a spherical cell of radius $a$. Here, $t$ is time and $(r,\theta, \phi)$ are spherical coordinates. It is convenient to introduce the Cartesian coordinates $(x,y,z)$, such that $r = \sqrt{x^2+y^2+z^2}$ is the radial distance from the cell center,  $\theta$ is the polar angle measured from the $z$-direction in the $(x,z)$ plane, and  $\phi$ is the azimuthal angle in the $(x,y)$ plane (Fig.~\ref{fig: pure_diff}A). It is also convenient to introduce  orthonormal frames $(\mathbf{e}_r,\mathbf{e}_\theta, \mathbf{e}_\phi)$ and $(\mathbf{e}_x,\mathbf{e}_y,\mathbf{e}_z)$ associated with the spherical and Cartesian coordinates with $\mathbf{e}_z = \cos\theta\mathbf{e}_r - \sin\theta \mathbf{e}_\theta$.

Considering diffusion of nutrients at constant diffusivity  $D$~\cite{berg1977physics} and starting from an initial concentration field $C(0, r,\theta, \phi) = C_b$, the unsteady diffusion equation subject to 
zero concentration at the cell surface and proper convergence to $C_b$ as $r\to\infty$ is given by,
\begin{align} \label{eq:diffusion}
\begin{split}
    \frac{\partial C}{\partial t} & = D \nabla^2 C, \\
    C|_{t=0} = C_b, \quad C|_{r=a} = 0, \quad &
    C|_{r\rightarrow\infty} = C_b \quad (\text{or} \;
     \nabla C|_{r\rightarrow\infty} = \nabla C_b).
\end{split}
\end{align}
If the background concentration is uniform $C(0, r,\theta, \phi)= C_b=C_o$, where $C_o$ is constant, the concentration field evolves as $C(t, r,\theta, \phi) = C_o[1 - a\erfc{\left((r-a)/\sqrt{4t}\right)}/r]$~\cite{carslaw1959conduction}.

Nutrient uptake is defined as the surface integral of the nutrient flux over the cell surface ${\rm I} = -\oint \mathbf{n}\cdot(-D\nabla C) dS$; here $\mathbf{n}$ is the outward unit normal, $d{\rm S} = 2\pi a^2 \sin\theta d\theta$ is the area element on the sphere, and the sign convention is chosen such that ${\rm I}$ is positive if the cell takes up nutrients,
\begin{align} \label{eq:I_flux}
\begin{split}
    {\rm I (t)} = \int_S D\nabla C\cdot \mathbf{n}dS 
    = 2\pi a^2D \int_{0}^{\pi} \frac{\partial C(t)}{\partial r} |_{r=a} \sin\theta d\theta.
\end{split}
\end{align}
In a uniform background concentration~\cite{karp1996, magar2003, michelin2011, michelin2013unsteady}, the nutrient uptake at the cell surface is given by ${\rm I}(t) = 4\pi a D C_o(1 + a/\sqrt{\pi t})$.  At steady state, it is  ${\rm I}(t\rightarrow\infty) = 4\pi a D C_o$.

In the following, we consider a background concentration $C_b = C_o(1 \pm z/L)$ that changes linearly along the $z$-axis from the base level $C_o$; here, $1/L$ represents the constant slope of the concentration gradient, taken such that the length scale $L \gg a$ is larger than the cell radius $a$. As $L \rightarrow \infty$, we recover the uniform concentration field. 

\paragraph{Non-dimensional equations}
Consider the length scale $a$, time scale $a^2/D$, and concentration scale $C_o$. Introduce the dimensionless concentration $c = C/C_o$, and redefine time $t := t/(a^2/D)$ and all spatial coordinates $(\cdot)\coloneqq (\cdot)/a$. Letting $\epsilon = a/L$ be the ratio between the cell size and concentration gradient,
the dimensionless counterparts to the unsteady diffusion equation and boundary conditions in~\eqref{eq:diffusion} are given by
\begin{align} \label{eq: unsteady_diffusion_eqs}
    \frac{\partial c}{\partial t} = \nabla^2 c,\quad c(t=0) = 1+\epsilon z, \quad c(r=1) = 0, \quad \nabla c(r\rightarrow\infty) = \epsilon \mathbf{e}_z.
\end{align}
The dimensionless counterpart to the nutrient uptake at the cell surface is written as
\begin{align} \label{eq:J_flux}
\begin{split}
    {\rm J (t)} = \dfrac{\rm I (t)}{4\pi a D C_o} =
     \frac{1}{2}\int_{0}^{\pi} \frac{\partial c(t)}{\partial r} |_{r=1} \sin\theta d\theta.
\end{split}
\end{align}
In a uniform background concentration~\cite{karp1996, magar2003, michelin2011, michelin2013unsteady}, the dimensionless nutrient uptake at the cell surface is given by ${\rm J}(t) = (1 + 1/\sqrt{4\pi t})$.  At steady state, it is   ${\rm J} = 1$.

\paragraph{Analytical solution}
To obtain the time evolution of the concentration field, we solved~\eqref{eq: unsteady_diffusion_eqs} analytically for $c(t,r,\theta)$; the $\phi$ coordinate is ignorable because of axisymmetry about the $z$-axis (Fig.~\ref{fig: pure_diff}A). Specifically, we introduced the nonlinear transformation $\mu = \cos\theta$ and used the Laplace transform to get the unsteady solution,
\begin{align}\label{eq:analytical_c}
\begin{split} 
    c(t, r, \mu) = & 1 - \frac{1}{r} \erfc{\left(\frac{r-1} {\sqrt{4t}}\right)} + \epsilon\mu \left(r-\frac{1}{r^2} \right)\\
    & -   \frac{2\epsilon\mu }{\pi r^2}\int_{0}^{\infty} e^{-u^2 t} \frac{u(r-1)\cos\left(u(1-r)\right) + (1 + ru^2)\sin\left(u(1-r)\right)}{u(1+u^2)} du.
\end{split}
\end{align}
Here, $u$ is a placeholder variable for integration; details to arrive at~\eqref{eq:analytical_c} are provided in~\ref{append1}. To obtain the steady-state solution $c(r,\mu)$, we take the limit $t\rightarrow \infty$,
\begin{align}
\begin{split}
    c(r, \mu) &= 1-\frac{1}{r}+\epsilon\mu \left(r-\frac{1}{r^2}\right).
\end{split}
\end{align} 
For $\epsilon=0$, the unsteady and state-state solutions are consistent with the classical solutions of the diffusion equation in uniform concentration \cite{carslaw1959conduction}.

\paragraph{Numerical solution} 
Next, we solved for the unsteady diffusion equation numerically using the Legendre spectral method in $\theta$ and employing a finite difference discretization in $r$ with a stretched mesh \cite{michelin2011,liu2024optimal, liu2024feeding}. For time stepping, we evolved the unsteady diffusion equation using the Backward Euler method (see details in \ref{append2}). 

To validate the numerical solution, we computed the spatiotemporal evolution of the concentration field $c$ for $\epsilon = [0, 0.01, 0.05]$ respectively. For each  $\epsilon$, we used the analytical solution~\eqref{eq:analytical_c} to compute the relative error $||c - c_{\rm analytical}||/||c_{\rm analytical}||$ and $||J - J_{\rm analytical}||/||J_{\rm analytical}||$ as a function of mesh size.  By constraining the relative error to be less than $10^{-4}$, we selected the appropriate spatial mesh size in $r$ and time step size.

\paragraph{Diffusive nutrient uptake in linear concentration gradients} In Fig.~\ref{fig: pure_diff}B, we show the time evolution of the concentration field in a uniform background concentration $\epsilon = 0$ (top row) and in a linear concentration gradient $\epsilon = 0.05$ (bottom row). 
By choice, the linearly changing background concentration $C_b$ is equal to $C_o$ at the cell center $z=0$; it increases linearly for positive $z$ and decreases by the same amount for negative $z$. As a result, the concentration field is asymmetric along the $z$-axis, compared to the uniform background concentration. But, by construction, this asymmetry leads to the same time evolution of the nutrient uptake $J(t)$ (Fig.~\ref{fig: pure_diff}C) because the nutrient flux on the positive and negative half-spheres balance each other. Indeed, from the property of Legendre polynomials $\int_{-1}^1 P_n(\mu)d\mu = 2\delta_{n0}$, the terms in~\eqref{eq:analytical_c} associated with concentration gradient $\epsilon$ do not contribute to the integral in~\eqref{eq:J_flux}. That is, although the spatial concentration fields differ between the uniform and linear gradient background (Fig.~\ref{fig: pure_diff}B,D), by construction, they both give rise to the same nutrient uptake at the cell surface (Fig.~\ref{fig: pure_diff}C). The difference between the two concentration fields is evident when examining their variation in the $z$-direction at steady state (Fig.~\ref{fig: pure_diff}D).

\section{Cilia-driven nutrient transport in linear concentration gradients}
\paragraph{Problem setting}
To generate feeding currents in a viscous fluid, a ciliated microorganism, through ciliary activity, executes a series of irreversible surface deformations. We consider sessile ciliates, that attach to a surface and don't swim.  A reaction force, typically supplied by a tether or a stalk, resists the net force due to ciliary activity,  causing it to remain in place. We ask, for a fixed rate of energy dissipation in the fluid, which ciliary motions maximize nutrient flux to the cell’s surface in a linear concentration gradient? And how nutrient uptake vary in linear gradients compared to uniform background concentration? To address these questions, we model the sessile ciliated cell using Blake's envelope model~\cite{blake1971, michelin2011, liu2024optimal, liu2024feeding}. The fluid in the three-dimensional domain internally bounded by the spherical cell is governed by incompressible Stokes equation,
\begin{align}\label{eq:stokes}
    -\nabla p + \eta\nabla^2\mathbf{u}=0,\qquad \nabla\cdot\mathbf{u}=0,
\end{align}
where $\mathbf{u}$ is the flow field, $p$ is the pressure field, and $\eta$ is the dynamic viscosity. The boundary conditions are given by,
\begin{equation}
    \left. \mathbf{u} \right|_{r = a} = \sum_{n=1}^\infty B_n V_n(\mu) \mathbf{e}_\theta, \qquad
    \left. \mathbf{u} \right|_{r \rightarrow \infty} = \mathbf{0},
\end{equation}
where $V_n(\mu) = \frac{2}{n(n+1)}\sqrt{1-\mu^2}P_n'(\mu)$ is velocity basis with $P_n(\mu)$ representing for the $n^{th}$ order of Legendre polynomial and $P_n'(\mu) = dP_n(\mu)/d\mu $. The constant coefficients $B_n$ represent the intensity of each velocity mode. The solution to Stokes equation~\eqref{eq:stokes} is given analytically in the form \cite{blake1971, liu2024optimal, liu2024feeding},
\begin{equation}
\begin{split} 
    u_{r}(r,\mu) &=  \sum_{n=1}^{\infty}\left(\dfrac{a^{n+2}}{r^{n+2}}-\dfrac{a^{n}}{r^{n}}\right)B_{n}P_{n}(\mu), \\ u_{\theta}(r,\mu) &= \sum_{n=1}^{\infty}\dfrac{1}{2}\left(\dfrac{na^{n+2}}{r^{n+2}}-\dfrac{(n-2)a^{n}}{r^{n}}\right)B_{n}V_{n}(\mu),\\
    p(r,\mu) &= p_\infty - \eta\sum_{n=1}^{\infty}\dfrac{4n-2}{n+1}\dfrac{a^{n}}{r^{n+1}}B_{n}P_{n}(\mu),
    \end{split}
\end{equation}
The hydrodynamic force acting on the sphere is obtained from $\mathbf{F}_h = \int \boldsymbol{\sigma}\cdot\mathbf{n}dS$, where $\boldsymbol{\sigma} = -p\mathbf{I}+\eta(\nabla \mathbf{u} + \nabla ^T\mathbf{u})$ is the stress tensor. The force required to keep the spherical cell attached is obtained from balance force, $\mathbf{F}_{\rm attach} = -\mathbf{F}_h = -4\pi\eta a B_1 \mathbf{e}_z$ \cite{liu2024feeding, liu2024optimal}. The energy dissipation rate is given by~\cite{kim1991, michelin2011, liu2024optimal} 
\begin{equation}
    \mathcal{P} = \int_{S} \mathbf{u}\cdot(\boldsymbol{\sigma}\cdot\hat{\mathbf{n}}) \textrm{d}S = 16\pi a\eta\sum_{n=1}^{\infty} \dfrac{B^2_n}{n(n+1)}.
\end{equation}

Coupled with the ciliate's induced flow field, the nutrient concentration field is governed by an advection-diffusion equation. With focusing on transient nutrient concentration evolution, we consider far-field boundary condition preserved within the considered time. The governing equation, initial condition and boundary conditions are
\begin{equation}
\begin{split}
    & \frac{\partial C}{\partial t} + \mathbf{u}\cdot \nabla C = D\nabla^2 C, \\
    \quad C|_{t=0} = C_b,& \quad C|_{r=a} = 0, \quad \nabla C|_{r\rightarrow\infty} = \epsilon \mathbf{e}_z.
    \label{eq:unsteady_ade}
\end{split}
\end{equation}

\paragraph{Non-dimensional equations}  Consider the length scale $a$, concentration scale $C_o$, and velocity scale $\mathcal{U}$ based on the energy dissipation rate $\mathcal{U} = \sqrt{\mathcal{P}/(8\pi a\eta)}$ \cite{liu2024optimal}. Here, we have two time scales: the advection time scale $\tau_{\rm advection} = a/\mathcal{U}$ and the diffusion time scale $\tau_{\rm diffusion} = a^2/D$. Choosing the advection time scale $\tau_{\rm advection}$, we obtain the dimensionless transport equation, together with the initial and boundary conditions,
\begin{align}
\begin{split}
\label{eq:adc_diff}
    &\frac{\partial c}{\partial t} + \mathbf{u}\cdot\nabla c = \frac{1}{\rm Pe}\nabla^2 c,\\
    c|_{t = 0} = 1 +& \epsilon z,\quad
    c|_{r=1} = 0,\quad
    \nabla c|_{r\rightarrow\infty} = \epsilon \mathbf{e}_z,
\end{split}
\end{align}
where the P\'{e}clet number ${\rm Pe} =  a\mathcal{U}/D$ is the ratio between the advection and diffusion rates. 

\paragraph{Decomposition of the concentration field}
To simplify the solution to the advection-diffusion equation in~\eqref{eq:adc_diff}, we introduce the concentration field  $c' = c-(1+\epsilon z)$, which represents the disturbance to the background concentration due to the presence of the cell. This field satisfies
\begin{align}
\begin{split} \label{eq: ade_disturbance}
    &\frac{\partial c'}{\partial t} + \mathbf{u}\cdot\nabla c'+\epsilon u_{z} = \frac{1}{\rm Pe}\nabla^2 c',\\
    c'|_{t = 0} = 0, &\qquad
    c'|_{r=1} = -\left(1+\epsilon\mu\right),\qquad
    \nabla c'|_{r\rightarrow\infty} = \mathbf{0},
\end{split}
\end{align}
where $u_z = \mathbf{u}\cdot\mathbf{e}_z$ is the fluid velocity component along $\mathbf{e}_z$ direction. By linearity of the advection-diffusion equation, we further decompose the concentration field $c'$ into two parts $c' = c_h+\epsilon \tilde{c} $, where $c_h$ is the solution associated with homogeneous ambient concentration background, and $\tilde{c}$ is the correction due to the concentration gradient. Substituting back into~\eqref{eq: ade_disturbance},  we get two sets of differential equations and boundary conditions. The equations governing $c_h(t,r,\theta)$ are given by
\begin{equation}
\begin{split}
    &\frac{\partial c_h}{\partial t} + \mathbf{u}\cdot\nabla c_h = \frac{1}{\rm Pe} \nabla^2 c_h, \\ 
    \quad c_h|_{t=0}  = 0,& \qquad
    c_h|_{r=1} = -1, \qquad \nabla c_h|_{r\rightarrow\infty} = \mathbf{0},\label{eq:ade_uniform}
\end{split}
\end{equation}
The equations governing $\tilde{c}(t,r,\theta)$ are given by
\begin{equation}
\begin{split}
    &\frac{\partial \tilde{c}}{\partial t} + \mathbf{u}\cdot\nabla \tilde{c}+ u_{z} = \frac{1}{\rm Pe}\nabla^2 \tilde{c},
    \\
     \tilde{c}|_{t=0}  = 0, &\qquad
    \tilde{c}_{r=1} = -\mu,\qquad \nabla\tilde{c}_{r\rightarrow\infty} = \mathbf{0}. \label{eq:ade_gradient}
\end{split}
\end{equation}
That is, the concentration $\tilde{c}(t,r,\mu)$ follows an advection-diffusion process with a distributed 'source' term represented by $u_z(r,\mu)$. This source term describes a supply of nutrients by the component $u_z(r,\mu)$ of cilia-induced flow that is aligned with the background concentration gradient in the $\mathbf{e}_z$ direction. The strength of this source term decays as $1/r$, indicating that the cell-induced effect decays inversely with distance from the cell. 

The dimensionless nutrient intake is given as the sum of nutrient intake $J_h$ in a uniform concentration field and nutrient intake $\tilde{J}$ in a concentration gradient field, calibrated to have zero background concentration at the cell center.
\begin{align}
\begin{split}
    J(t) & 
    = J_{h}(t) + \epsilon\tilde{J}(t) = \frac{1}{2}\int_{-1}^1 \frac{\partial c_h}{\partial r}\Big|_{r=1} d\mu + \epsilon\frac{1}{2}\int_{-1}^1 \frac{\partial \tilde{c}}{\partial r}\Big|_{r=1} d\mu  .\label{dimensionlessJ}
    \end{split}
\end{align}

\begin{figure*}[tbhp!]
\centerline{\includegraphics{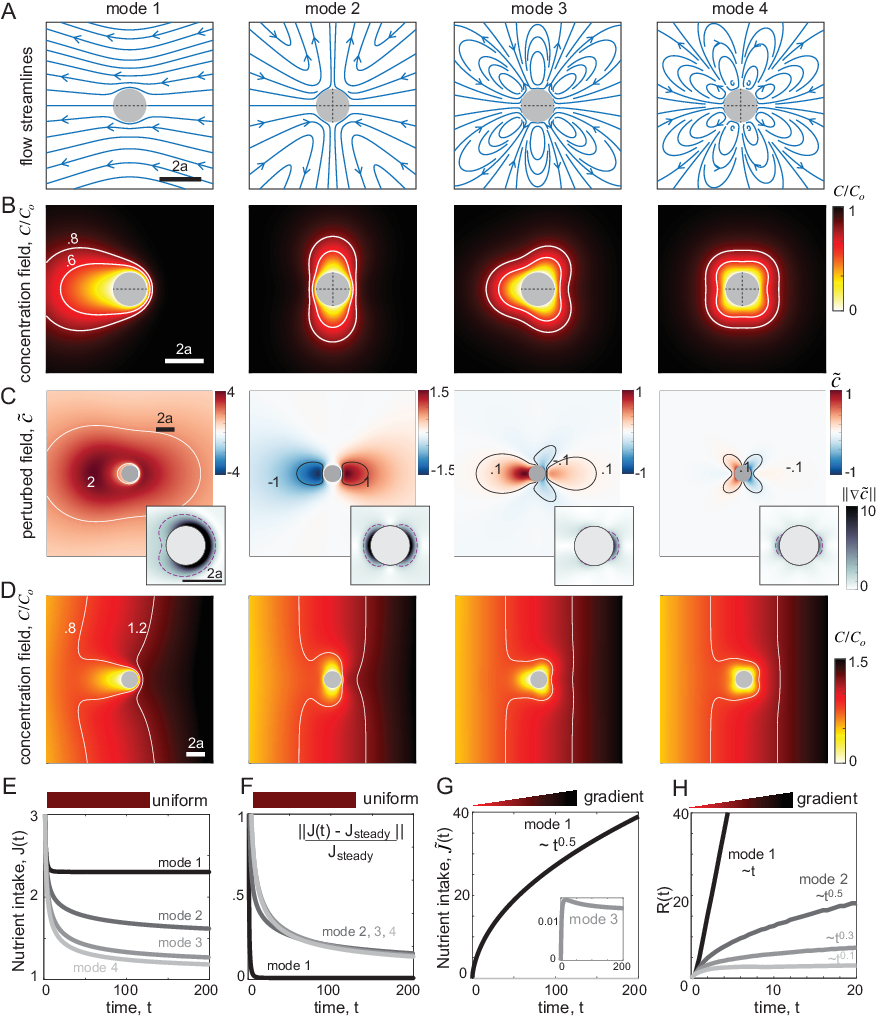}}
    \caption[]{\textbf{Unsteady transport in uniform and linear concentration.} A. Flow streamlines around a sessile ciliated cell generated by the first four surface velocity modes. Concentration field $c$ in B. uniform and D. linear gradient, where $c=c_h + \epsilon \tilde{c} + (1+\epsilon z)$. C.  snapshots of $\tilde{c}$ with insets showing $|\nabla \tilde{c}|$ around the cell. E. Time evolution of nutrient intake $J_h(t)$ in uniform concentration. 
    In uniform background concentration, E. mode 1 exhibits the largest nutrient intake and F. fastest convergence to steady-state. In linear gradients, G. the component $\tilde{J}(t)$ is identically zero for even modes and largest for the first mode. H. the distance $R(t)$ where $\tilde{c}$ begins to reach $10\%$ grows with time.
     }
\label{fig: concentration_c_t}
\end{figure*}

\paragraph{Unsteady transport in uniform background concentration}
Consider a sessile ciliated sphere generating feeding currents in a uniform concentration with $\epsilon = 0$. The steady-state limit of this problem has been studied~\cite{purcell1977life, karp1996, magar2003, michelin2011, kanso2021teamwork, liu2024optimal,liu2024feeding}. Here, we focus on the unsteady dynamics of nutrient transport. By definition, the concentration field is $c(t, r,\mu) = 1+ c_h(t,r,\mu)$.
We numerically solved equation ~\eqref{eq:ade_uniform} for $c_h(t,r,\mu)$ (\ref{append2}).
In Fig.~\ref{fig: concentration_c_t}A, we show the streamlines (blue) of flow fields generated by the ciliated sphere with surface velocity corresponding to one of the first four modes: 'treadmill' (mode 1), dipolar (mode 2), tripolar (mode 3), and quadripolar (mode 4). In all cases, we constrained the total hydrodynamic power $\mathcal{P}/8\pi\eta a = 1$ to be constant. Only the treadmill mode $(n=1)$ contributes a net hydrodynamic force that needs to be balanced by an attachment force. By symmetry, the subsequent three higher-order modes produce zero force.  Notice that odd modes ($n=1,3$) produce flows with lateral (left-right) symmetry while even modes ($n=2,4$) result in flows with both lateral and front-back symmetry.

We solved~\eqref{eq:ade_uniform} at Pe $= 10$. Snapshots of the concentration field $c(t, r,\mu)$ for each mode at time $t=20$ are shown in Fig.~\ref{fig: concentration_c_t}B. The concentration fields reflect the symmetries of the underlying flows. We next calculated the time evolution of the nutrient intake $J(t)$ corresponding to each of these four modes (Fig.~\ref{fig: concentration_c_t}E). Compared to higher-order modes, the treadmill mode, at the same hydrodynamic power,  provides the largest nutrient intake to the cell,  consistent with findings in \cite{michelin2011, liu2024optimal} at steady-state. Importantly, depending on the surface mode, the rate of change of nutrient intake approaches steady-state at different rates: the treadmill mode approaches steady solution at a faster rate compared to higher-order modes (Fig.~\ref{fig: concentration_c_t}F). 

\paragraph{Unsteady transport in linear gradients}
Consider now the sessile ciliated sphere in a linear concentration gradient with $\epsilon = 0.05$. 
Given our numerical solution $c_h(t,r,\mu)$ of~\eqref{eq:ade_uniform},  we separately solved \eqref{eq:ade_gradient} for $\tilde{c}(t,r,\mu)$.  We substituted the numerical solutions into $c = c_h + \epsilon \tilde{c} + (1+\epsilon z)$ to recover the full concentration field. 
The concentration field $\tilde{c}$ is shown in Fig.~\ref{fig: concentration_c_t}C for all four surface velocity modes. 
The insets show a zoom in on the concentration gradient $|\nabla \tilde{c}|$ around the cell, with dashed contour lines $|\nabla \tilde{c}|=2$ shown to highlight the fact that these contour lines are front-back asymmetric for odd modes and front-back symmetric for even modes -- with opposite sign of the concentration gradient. 
Fig.~\ref{fig: concentration_c_t}G shows the corresponding nutrient intake $\tilde{J}(t)$ for each mode. 
As $\tilde{c}(t,r,\mu)$ changes in time, the nutrient intake $\tilde{J}(t)$ associated with the even modes is identically zero for all time because of the front-back flow symmetries of these modes;  $\tilde{J}(t)$ associated with the first mode increases with the square root of  time, but $\tilde{J}(t)$ associated with the third mode converges to a constant as time increases.

To emphasize the role of the 'forcing term' $u_z(r,\mu)$ due to cilia-induced flows in driving nutrients to the cell surface  in~\eqref{eq:ade_gradient},  we introduced a distance $R(t)$ that captures how, starting from an identically zero initial value $\tilde{c}(t=0,r,\mu)=0$, $u_z(r,\mu)$ creates a region of high concentration around the cell. Specifically, we defined $R(t)$ such that $\{ \tilde{c}<0.1 \; \text{for} \; r>R \}$; the distance $R(t)$ marks the region within which $\tilde{c}$ increases from zero up to a percentage of the normalized baseline $C/C_o = 1$, beyond $R(t)$, this concentration drops to below $10\%$. 
This distance can be used for approximating the growth domain of the concentration field $\tilde{c}(t,r,\mu)$.
The time evolution of this distance $R(t)$ is shown in Fig.~\ref{fig: concentration_c_t}H; $R$ scales with time as follows: for mode 1, $R\sim t$ grows linearly in time; 
for mode 2, $R\sim \sqrt{t}$ grows proportionally to the square root of time, for modes 3 and 4, $R\sim t^{0.3}$ and $R\sim t^{0.1}$, respectively, indicating, as expected, that $R$ decreases with higher-order modes. Mode 1 maximizes transport to the cell surface.

\begin{figure*}[tbhp!]
	\centerline{\includegraphics{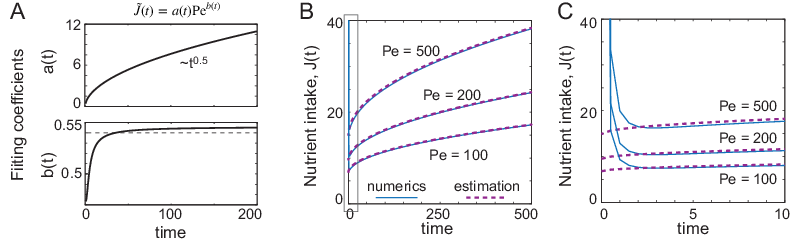}}
	\caption[]{\textbf{Nutrient intake of the first mode.} A. Coefficients of curve fitting of nutrient intake $\tilde{J}(t)$ corresponding to mode 1 versus Pe at every instant time. B. Nutrient intake comparison between numerical computation $J(t) = J_h(t) + \epsilon \tilde{J}(t)$ and estimation $J(t) = J_h|_{t\rightarrow\infty} + 0.94 \epsilon \sqrt{t\;{\rm Pe}}$ under Pe $= [100, 200, 500]$ with time range $t\in[0,500]$. C. Zoom-in version for nutrient intake comparison for time range $t\in [0,10]$. All calculations are with gradient $\epsilon = 0.05$. }
    \label{fig: Jtilde_mode1}
\end{figure*}

We next computed the concentration field $c(t,r,\mu)$ and nutrient intake $J(t,r,\mu)$ for mode 1 for a range of P\'{e}clet number ${\rm Pe}\in[1,100]$ and a range of time $t\in[0,200]$. 
We fitted the component $\tilde{J}$ of the nutrient intake using a power law model $\tilde{J}= a(t) {\rm Pe}^{b(t)}$. The fitting coefficients $a(t)$ and $b(t)$ are shown in Fig.~\ref{fig: Jtilde_mode1}A, where at all time the R-squared values are larger than $0.99$: $b(t)$ converges to ${0.54}$ after time $t> 50$ and $a(t)$ scales with $t^{0.5}$ at all time, giving rise to $\tilde{J}\sim t^{0.5} {\rm Pe}^{0.54}$. 
We thus simplified the scaling law $\tilde{J}(t) = c(\sqrt{t \rm{Pe}})^d$ and found that $\tilde{J}(t) = 0.94(t\;{\rm Pe})^{0.55}$, with an $R$-squared value larger than $0.99$, which we approximated further to $\tilde{J}(t) = 0.94 \sqrt{t\;{\rm Pe}}$.
Substituting back into~\eqref{dimensionlessJ}, we got an approximate expression for the optimal nutrient intake $J(t)$ associated with mode 1 in linear concentration gradients,
\begin{equation}
    J(t) = J_h(t) + 0.94\epsilon\sqrt{t\;{\rm Pe}}.
    \label{eq:Jmodel}
\end{equation}
In Fig.~\ref{fig: Jtilde_mode1}B, we tested the ability of this power law to predict nutrient uptake beyond the ranges of P\'{e}clet number ${\rm Pe}\in[1,100]$ and time $t\in[0,200]$ that were used to obtain the fitted coefficient. In particular, we compared the model prediction in~\eqref{eq:Jmodel} to results from full numerical simulations for Pe = 100, 200, and 500, for a time range extending up to $t = 500$. The agreement between the model in~\eqref{eq:Jmodel} and the numerical simulations beyond a short time range is excellent. To interrogate the initial mismatch at short time scales, see the zoom in version in Fig.~\ref{fig: Jtilde_mode1}C; again, beyond an initial time window not exceeding $t=5$, the model exhibits remarkable agreement with simulations. 

\begin{figure*}[tbhp!]
	\centerline{\includegraphics{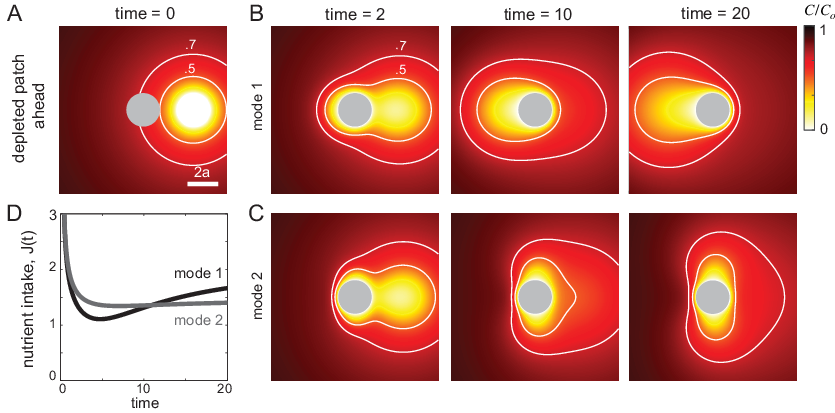}}
	\caption[]{\textbf{Nutrient uptake in a concentration field with depleted upstream patch.} A. The initial depleted patchy concentration upstream of the model cell. B-C. The concentration field snapshots at time $t = [0,2,10,20]$ corresponding to surface velocity with B. mode 1 and C. mode 2. D. The nutrient intake $J(t)$ versus time corresponds with input surface velocity with only mode 1 (black) and mode 2 (grey). All case studies are under Pe $= 10$. }
    \label{fig: c_patches}
\end{figure*}

\begin{figure*}[tbhp!]
	\centerline{\includegraphics{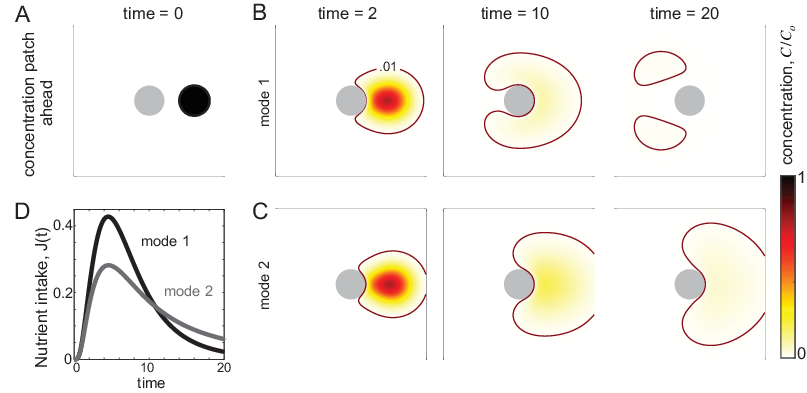}}
	\caption[]{\textbf{Nutrient uptake in a depleted environment with a rich concentration patch.}  A. The initial field contains a concentrated blob upstream of the model cell. B-C. The concentration snapshots at time $t = [0,2,10,20]$ corresponding to surface velocity with B. mode 1 and C. mode 2. D. Nutrient intake versus time corresponds to surface velocity with mode 1 (black) and mode 2(grey). All case studies are under Pe $= 10$. }
    \label{fig: c_blob}
\end{figure*}

\paragraph{Unsteady transport in patchy concentration fields}
Lastly, we considered two case studies that represent the feeding of sessile ciliate in a patchy concentration background.  
First, consider an initial concentration field $c(0,r,\mu) = 1 - R_{b}/|\mathbf{r}-\mathbf{r}_{0,b}|$ that is a steady-state solution of the diffusion equation outside an empty blob of size $R_b$ centered at $r_{0,b}$. In Fig.~\ref{fig: c_patches}A, we show this initial concentration field for $R_b = 1$ and $\mathbf{r}_{0,b} = 3\mathbf{e}_z$, representing a depleted concentration patch upstream of the mode ciliate. Using this initial concentration field, we directly solved the unsteady state advection-diffusion equation~\ref{eq:unsteady_ade} numerically at Pe $= 10$ for mode 1 and mode 2 at the same constant energy dissipation rate, as done before. The time evolution of the concentration field associated with each mode is shown in Fig.~\ref{fig: c_patches}B, C for mode 1 and mode 2, respectively. The corresponding nutrient intake $J(t)$ is shown in Fig.~\ref{fig: c_patches}D. These results show that mode 2 outperforms mode 1 at an initial transient time, but mode 1 takes over as the optimal mode in the long term.
We next considered an initial concentration field, where the entire domain is depleted except for a nutrient-rich blob (shown in black in Fig.~\ref{fig: c_blob}A) placed on one side of the ciliated sphere. The time evolution of this nutrient-rich patch due to cilia-induced flows is shown in Fig.~\ref{fig: c_blob}B, C for mode 1 and mode 2, and the nutrient intake versus time is shown in Fig.~\ref{fig: c_blob}D. In contrast to the depleted nutrient patch study, mode 1 is optimal during the initial transient phase, while mode 2 becomes more advantageous over the long term. 

Taken together, these results show that in a linear concentration gradient with potentially infinite nutrient supply, mode 1 is optimal when the direction of cilia-induced velocity is aligned with the direction of the nutrient concentration gradient. However, in nutrient-depleted domains with sparse nutrient availability, mode 1 may become suboptimal. An adaptive adjustment of beating modes needs to be considered for optimizing nutrient intake in non-uniform fields.
These findings underscore that in dynamically changing environments, there is no unique strategy or surface activity that maximizes uptake at all time.

\section{Summary}

We proposed a mathematical model for sessile ciliates generating feeding currents in concentration gradients. In still water, in the absence of feeding currents, the problem is governed by nutrient diffusion only. In this diffusive limit, starting from an initial linear concentration field, we provided an analytical solution for the unsteady concentration field around a spherical cell. By symmetry, the diffusion process in such linear gradients provides no benefits over what is achievable by a sessile cell in a uniform background concentration. 

Given the ability of the ciliated cell to generate feeding currents and considering ciliary surface activities that are axisymmetric about the direction of concentration gradient, we employed Blake's envelope model to solve for the feeding currents in terms of Legendre polynomials~\cite{blake1971,magar2003,michelin2011,liu2024feeding,liu2024optimal}. We considered surface activity associated with the first four Legendre modes. To properly compare feeding efficiencies across different modes, we fixed the hydrodynamics dissipation power to be constant. We then formulated the feeding problem as an advection-diffusion process in linear concentration gradients.
We decomposed the concentration field into two fields, governed, respectively, by a homogeneous and a "forced" advection-diffusion process. We solved for each field separately using a spectral method~\cite{michelin2011, michelin2013unsteady}. We found that, in  linear concentration gradients, only the odd modes contribute to nutrient uptake. Odd modes generate asymmetric flows along the direction of concentration gradient that increase nutrient intake;  even modes generate symmetric flows and thus provide no advantage beyond what is achievable in a uniform concentration field. 
Moreover, only the treadmill surface activity (mode 1) contributes significantly to sustained nutrient intake, while higher-order modes have only a minor impact on nutrient intake.
Focusing on the treadmill mode and analyzing how nutrient uptake scales with Péclet and time, we obtained an analytic expression $J(t) = J_h|_{t\rightarrow\infty} + 0.94\epsilon \sqrt{t\; {\rm Pe}}$, where $\epsilon$ is the concentration gradient. This approximate model accurately predicts the optimal rate of nutrient intake in the long time regime and large P\'{e}clet for a sessile ciliate in linear concentration gradients. Lastly, we considered nutrient intake in patchy concentration fields. Considering both depleted and nutrient-rich patches upstream of the cell, we found that shifting between ciliary beating modes optimizes cumulative nutrient intake beyond what is achievable by a single mode.

Our findings demonstrate that directional cilia beating aligned with the concentration gradient is essential for maximizing feeding performance. As such, they offer new perspectives on the mechanisms by which sessile ciliates could regulate their orientation in response to concentration gradients, say relative to the substrate to which they are attached~\cite{pepper2013, pepper2021effect}, to maximize feeding flux. Extending our approach to motile ciliates in concentration gradients would open up exciting opportunities to study chemotaxis~\cite{goldstein2015green,kirkegaard2018role, nelson2023cells, zhu2022optimizing, fang2023data}.
More generally, our findings are important in predicting the ecological impact of sessile ciliates on both local and global nutrient distributions within microbial communities \cite{dopheide2009relative, sherr2002significance}.


\printbibliography

\null
\vfill
\newpage

\renewcommand\thesection{APPENDIX \Alph{section}}
\setcounter{section}{0}

\section{Diffusion in concentration gradients}
\label{append1}
We consider the concentration field $c' = c-(1+\epsilon z)$. Substitute back into~\eqref{eq: unsteady_diffusion_eqs}, using the nonlinear transformation  $\mu = \cos\theta$, the concentration field $c'$ satisfies
\begin{equation} \label{eq: unsteady_diffusion_c_prime}
\begin{split} 
    \frac{\partial c'}{\partial t} & =  \nabla^2 c', \qquad 
    \begin{cases}
        c(t=0, r, \mu) = 0, \\
        c(t, r=1, \mu) = -1 - \epsilon\mu,\\
        \nabla c(t, r\rightarrow\infty,\mu) = 0,
    \end{cases} 
\end{split}
\end{equation}
Take the Laplace transform of $c'$ to be $\bar{c}(s,r,\mu) \coloneqq \mathcal{L}\{c'\}(s,r,\mu) = \int_0^\infty c'(t,r,\mu)e^{-st}dt$; apply the Laplace transform to the governing equation and boundary conditions  to get
\begin{align}
    s\bar{c} & = \frac{\partial^2 \bar{c}}{\partial r^2} + \frac{2}{r}\frac{\partial \bar{c}}{\partial r} + \frac{1}{r^2}\frac{\partial}{\partial\mu}\Big[(1-\mu^2)\frac{\partial\bar{c}}{\partial\mu}\Big],\label{eq: laplace_diffusion} \\
   & \begin{cases} \bar{c}(s, r=1, \mu) = -\dfrac{1}{s}(1 + \epsilon\mu), \\
    \bar{c}(s, r\rightarrow\infty,\mu) = 0,  
    \end{cases} \label{eq: laplace_bc}
\end{align}
Use separation of variables $\bar{c} = \Phi(\mu)R(s,r)$ and substitute into~\eqref{eq: laplace_diffusion},
\begin{align}
     \frac{\partial^2 R}{\partial r^2}\Phi + \frac{2}{r}\frac{\partial R}{\partial r}\Phi + \frac{1}{r^2}\frac{\partial}{\partial\mu}\Big[(1-\mu^2)\frac{\partial\Phi}{\partial\mu}\Big]R - s\Phi R = 0.
\end{align}
Multiply by $\dfrac{r^2}{\Phi(\mu)R(s, r)}$, take all terms that depend on $\mu$ to one side, and define the integer number $n\geq 0$,
\begin{align}
     r^2\frac{d^2 R}{d r^2}\frac{1}{R} + 2r\frac{d R}{d r}\frac{1}{R}  - s r^2 = - \frac{d}{d\mu}\Big[(1-\mu^2)\frac{d\Phi}{d\mu}\Big]\frac{1}{\Phi} = n(n+1).
\end{align}
We arrive at two differential equations governing  $R(s, r)$ and $\Phi(\mu)$ separately
\begin{align}
    r^2\frac{d^2 R}{d r^2} + 2r\frac{d R}{d r}  - \left(s r^2 + n(n+1)\right)R = 0,\\
    \frac{d}{d\mu}\Big[(1-\mu^2)\frac{d\Phi}{d\mu}\Big] + n(n+1)\Phi = 0.
\end{align}
The function $\Phi(\mu)$ satisfies the Legendre differential equation, whose solution is given by the Legendre polynomials $P_n(\mu)$; the function $R(s, r)\equiv R(\sqrt{s}r)$ satisfies the modified spherical Bessel equation with $\sqrt{s}r$ as the independent variable, whose solution is given by the $n^{th}$ modified spherical Bessel functions of $\sqrt{s}r$. 
The terms related to modified Bessel function of the first kind $i_n(\cdot)$ are dropped to ensure convergence of the solution in the three-dimensional domain of interest is infinite, which extends to infinity and is bounded internally by a sphere. 
Putting the two solutions together, we get
\begin{align}\label{eq:laplace_general_solution}
    \bar{c}(s, r,\mu) = \sum_{n=0}^{\infty} A_n k_{n} (\sqrt{s}\; r)P_n(\mu),
\end{align}
where $A_n$ are constant coefficients, $P_n(\mu)$ the $n^{th}$ Legendre polynomial, $k_n(\cdot)$ the $n^{th}$ modified spherical Bessel function of the second kind.

Substitute the general solution in~\eqref{eq:laplace_general_solution} back into the boundary conditions in~\eqref{eq: laplace_bc},
\begin{align}\label{eq:bc_bessel}
        -\frac{1}{s} = A_0 k_{0}(\sqrt{s}),\qquad
        -\frac{\epsilon}{s} = A_1 k_{1}(\sqrt{s}),\qquad
        A_n = 0 \ \text{for} \  n\geq 2.
\end{align}
The first two modified Bessel functions of the second kind are
\begin{align}\label{eq:two_modified_bessel}
    k_0(z) = \frac{e^{-z}}{z},\quad k_1(z) = \frac{e^{-z}(z + 1)}{z^2}.
\end{align}
Solve for the unknown coefficients in~\eqref{eq:bc_bessel} and substitute back into~\ref{eq:laplace_general_solution},
\begin{align}
    \bar{c}(s,r,\mu) = -\frac{k_{0}(\sqrt{s}\; r)}{s k_{0}(\sqrt{s})}P_0(\mu) - \frac{\epsilon k_{1}(\sqrt{s}\; r)}{s  k_{1}(\sqrt{s})}P_1(\mu).
\end{align}
Use the Inversion Theorem to obtain the concentration field
\begin{align}\label{eq:inverse_laplace}
    c'(t,r,\mu) = -\frac{1}{2\pi i}\int_{\gamma - i\infty}^{\gamma+i\infty}  \left( \frac{k_{0}(\sqrt s \;r)}{s  k_{0}(\sqrt s)}P_0(\mu) + \frac{k_{1}(\sqrt s \; r)}{s k_{1}(\sqrt s )}\epsilon P_1(\mu) \right) e^{s t} ds.
\end{align}

\renewcommand\thefigure{A.\arabic{figure}}
\setcounter{figure}{0}

\begin{figure*}[tbhp!]	\centerline{\includegraphics{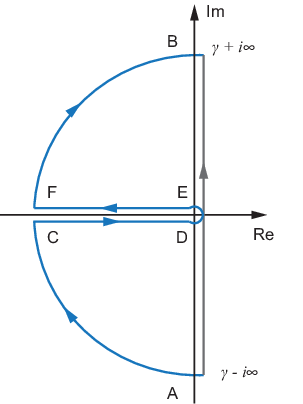}}
	\caption[]{\textbf{Integration path in the complex plane.} 
  Because the origin is a singular point of the integrand in~\eqref{eq:inverse_laplace}, the path of integration from $\gamma-i\infty$ to $\gamma+i\infty$ along the grey line AB is replaced with integration along the half circle in left plane~\cite{carslaw1959conduction}; specifically, the integral path is changed from the grey line AB to the blue line A-CDEF-B. Lines CD and EF are very close to the real axis, and the singular point at origin is avoided by the small circle DE.}
    \label{integral_path}
\end{figure*}

The integration path from $\gamma-i\infty$ to $\gamma+i\infty$ can be replaced by the integral along a half circle with infinite radius in the left half domain, along paths ACDEFB shown in Fig.~\ref{integral_path}, \cite{carslaw1959conduction}. Note that integral along the small circle DE of any function $f(z)$ is $\left( 2\pi i\; {\rm Res}_{s=0} f(z)\right)$, where {\rm Res} is the residual. Using~\eqref{eq:two_modified_bessel}, the residuals of the two terms in the integral~\eqref{eq:inverse_laplace} are given by 
\begin{align} \label{eq:residual}
    \begin{split}
        {\rm Res}_{s=0} \left( e^{st}\frac{k_{0}(\sqrt s \;r)}{s  k_{0}(\sqrt s)}\right) &= \lim_{s\rightarrow 0} \left( e^{st}\frac{k_{0}(\sqrt s \;r)}{ k_{0}(\sqrt s)}\right) = \frac{1}{r},\\
        {\rm Res}_{s=0} \left( e^{st}\frac{k_{1}(\sqrt s \;r)}{s  k_{1}(\sqrt s)} \right) &= \lim_{s\rightarrow 0} \left( e^{st}\frac{k_{1}(\sqrt s \;r)}{ k_{1}(\sqrt s)}\right) = \frac{1}{r^2}.
    \end{split}
\end{align}
Substituting the residuals~\eqref{eq:residual} into the integration~\eqref{eq:inverse_laplace}, we get
\begin{align}
\begin{split}
    c' &= - \frac{1}{2\pi i}\left[ {\rm I} \;P_0(\mu) + {\rm II} \;\epsilon P_1(\mu) + 2\pi i\left(\frac{1}{r}P_0(\mu) + \frac{1}{r^2}\epsilon P_1(\mu) \right) \right],\\
    &= - \frac{1}{2\pi i} \bigg( {\rm I}\; P_0(\mu) +  {\rm II}\; \epsilon P_1(\mu) \bigg) - \frac{1}{r}-\epsilon\mu\frac{1}{r^2},
\end{split}
\end{align}
where
\begin{align}\label{eq:integrals}
    {\rm I} = \int_{\rm CD+EF} \frac{k_{0}(\sqrt s \;r)}{s  k_{0}(\sqrt s)}\;ds,\qquad
    {\rm II} = \int_{\rm CD+EF} \frac{k_{1}(\sqrt s \;r)}{s  k_{1}(\sqrt s)}\;ds.
\end{align}
Considering $s = u^2e^{i\pi}$ for the integration in the upper left plane \cite{carslaw1959conduction}, we have the integral along line EF of the first term 
\begin{align}\label{eq:I_EF}
    {\rm I}_{\rm EF} = \int_{0}^{\infty}   e^{-u^2 t} \frac{k_{0}(ru e^{i\pi/2})}{ k_{0}(u e^{i\pi/2})}\frac{2du}{u}.
\end{align}
The relation of the modified spherical Bessel functions and the spherical Hankel functions are given by \cite{abramowitz1968handbook}
\begin{align}
\begin{split}
    k_n(x) = \begin{cases} -\frac{1}{2}\pi i^{n}h_n^{(1)}(ix),& -\pi<\arg x\leq\frac{\pi}{2},\\
    -\frac{1}{2}\pi i^{-n}h_n^{(2)}(-ix),& -\frac{\pi}{2}<\arg x\leq \pi,\end{cases}
    \label{eq:kn}
\end{split}
\end{align}
where $h^{(1)}_n$ and $h_n^{(2)}$ are spherical Hankel functions of the first and second kind; their relations to spherical Bessel functions $j_n$ and $y_n$ are given by
\begin{align}
\begin{split}
    h_n^{(1)}(x) &= j_n(x) + iy_n(x),\\
    h_n^{(2)}(x) &= j_n(x) - iy_n(x).
    \label{eq:sphericalHankel}
\end{split}
\end{align}
As a side note, the spherical Bessel and Hankel functions are related to the Hankel and Bessel functions $J_n, Y_n, K_n, I_n, H_n^{(1)}, H_n^{(2)}$ via $j_n(z) = \sqrt{\frac{\pi}{2z}}J_{n+1/2}(z)$, $y_n(z) = \sqrt{\frac{\pi}{2z}}Y_{n+1/2}(z)$, $h_n^{(1)}(z) = \sqrt{\frac{\pi}{2z}}H_{n+1/2}^{(1)}(z)$, $h_n^{(2)}(z) = \sqrt{\frac{\pi}{2z}}H_{n+1/2}^{(2)}(z)$.
Using the relations in~\eqref{eq:kn} and~\eqref{eq:sphericalHankel}, the integration in~\eqref{eq:I_EF} can be written as
\begin{align}
    {\rm I}_{EF} = \int_{0}^{\infty} e^{-u^2 t} \frac{h_0^{(2)}(ur )}{ h_0^{(2)}(u )}\frac{2du}{u}
    = \int_{0}^{\infty} e^{-u^2 t} \frac{j_0(ur ) - i y_0(ur )}{ j_0(u) - i y_0(u)}\frac{2du}{u}.
\end{align}
For the integration along CD in the lower left plane, we set $s = u^2e^{-i\pi}$. By similar procedure, we obtain 
\begin{align}
    {\rm I}_{\rm CD} = \int_{\infty}^{0} e^{-u^2 t} \frac{k_{0}(ur e^{-i\pi/2})}{ k_{0}(u e^{-i\pi/2})}\frac{2du}{u}
    = -\int_{0}^{\infty} e^{-u^2 t} \frac{j_0(ur ) + i y_0(ur )}{j_0(u) + i y_0(u)}\frac{2du}{u}.
\end{align}
Put together, the first integral in~\eqref {eq:integrals} is given by
\begin{align}
    {\rm I} = {\rm I}_{EF} + {\rm I}_{CD}
    = 4i \int_{0}^{\infty} e^{-u^2 t} \frac{j_0(ur )y_0(u) - y_0(ur)j_0(u)}{ j_0^2(u) +  y_0^2(u)}\frac{du}{u}.
\end{align}
Similarly, the second integral in~\eqref {eq:integrals} can be obtained
\begin{align}
    {\rm II} = 4i \int_{0}^{\infty} e^{-u^2 t} \frac{j_1(ur )y_1(u) - y_1(ur)j_1(u)}{ j_1^2(u) +  y_1^2(u)}\frac{du}{u}.
\end{align}
Putting all the pieces together, we get the solution of the concentration field
\begin{align}
\begin{split}
    c' =  - \frac{1}{r}-\frac{\epsilon\mu}{r^2} -\frac{2}{\pi} \int_{0}^{\infty} e^{-u^2 t} &\bigg(\frac{j_0(ur )y_0(u) - y_0(ur)j_0(u)}{ j_0^2(u) +  y_0^2(u)}\;P_0(\mu) + \\
    & \frac{j_1(ur )y_1(u) - y_1(ur)j_1(u)}{ j_1^2(u) +  y_1^2(u)} \; \epsilon P_1(\mu)\bigg)\frac{du}{u}.
    \end{split}
\end{align}
The actual concentration field $c = c' + 1+\epsilon z$ is given by 
\begin{align}
\begin{split}
    c = 1 - \frac{1}{r} + \epsilon\mu (r-\frac{1}{r^2}) - \frac{2}{\pi} \int_{0}^{\infty} e^{-u^2 t} &\bigg( \frac{j_0(ur )y_0(u) - y_0(ur)j_0(u)}{ j_0^2(u) +  y_0^2(u)}\;P_0(\mu) + \\
    &\frac{j_1(ur )y_1(u) - y_1(ur)j_1(u)}{ j_1^2(u) +  y_1^2(u)} \; \epsilon P_1(\mu)\bigg)\frac{du}{u}.
\end{split}
\end{align}
Alternatively, by rearranging terms, the solution can be written as
\begin{align}
\begin{split}                                
    c(t,r,& \mu) = 1 - \frac{1}{r}  \erfc{\left(\frac{r-1}{\sqrt{4t}}\right)}
    + \epsilon\mu \left(r-\frac{1}{r^2} \right)\\
    &+ \epsilon\mu \left( - \frac{2}{\pi r^2}\int_{0}^{\infty} e^{-u^2 t} \frac{u(r-1)\cos\left(u(1-r)\right) + (1 + ru^2)\sin\left(u(1-r)\right)}{u(1+u^2)} du \right).
\end{split}
\end{align}

\section{Spectral method for unsteady diffusion}
\label{append2}

We use a spectral method based to numerically solve~\ref{eq: unsteady_diffusion_c_prime}  for the unsteady concentration field $c'= c(t,r,\mu)-(1+\epsilon z)$.
Specifically, following \cite{michelin2011, liu2024optimal, liu2024feeding}, we expand the solution in terms of Legendre polynomials  
\begin{align}
    c'(t,r,\mu) &= \sum_{m=0}^\infty C_m(t,r)P_m(\mu).
\end{align}
We substitute this expansion into \eqref{eq: unsteady_diffusion_c_prime}, and project on the $k^{th}$ Legendre polynomial. Using properties of Legendre polynomials, we get
\begin{equation}
    \begin{split}
        &\frac{\partial C_k(t,r)}{\partial t} = \frac{1}{r^2} \left[ \frac{\partial}{\partial r} \Big(r^2\frac{\partial C_k(t,r)}{\partial r}\Big) - n(n+1)C_k(t,r) \right],\\
        &C_k|_{t=0} = 0,\\
        &C_k|_{r=1} = -1\delta_{0k} -\epsilon\delta_{1k},\\
        &\frac{dC_k}{dr}]\Big|_{r\rightarrow\infty} = 0.
    \end{split}
\end{equation}
Only the first two terms
$C_0(t,r)$ and $C_1(t,r)$ survive.
We discrete the space $r$  using a finite difference method and time $t$  using a Backward Euler method for $C_0(t,r)$ and $C_1(t,r)$. Finally, given $c'$ in terms of $C_0(t,r)$ and $C_1(t,r)$, we add the linear concentration background $ 1+\epsilon z$ to recover  the full concentration field $c(t,r, \mu)$.

We use a similar spectral method to solve the advection-diffusion equations in ~\eqref{eq:ade_uniform} and \eqref{eq:ade_gradient}. Namely,  we project $c_h$ and $\tilde{c}$ onto the Legendre polynomials $P_m(\mu)$ to obtain the expansions
\begin{align}
    c_h(r,\mu,t) &= \sum_{m=0}^\infty C_m(r,t)P_m(\mu),\label{eq:uniformexpand}\\
    \tilde{c}(r,\mu,t) &= \sum_{m=0}^\infty \tilde{C}_m(r,t)P_m(\mu).\label{eq:gradientexpand}
\end{align}
Substitute~\eqref{eq:uniformexpand}  into~\eqref{eq:ade_uniform}  to get
\begin{align} 
\begin{split} \label{eq:spectral_c_k_uniform}
   \frac{\partial C_k}{\partial t} + \sum_{n=1}^{\infty}\sum_{m=0}^{\infty}B_{n} &\left(E_{mnk}f_{nr}\frac{\partial C_{m}}{\partial r} - F_{mnk}f_{n\theta} \frac{C_{m}}{r}\right) \\
    & = \frac{1}{\rm Pe}\left(\frac{\partial^{2}C_{k}}{\partial r^{2}}+ \frac{2}{r}\frac{\partial C_{k}}{\partial r} - \frac{k(k+1)}{r^2}C_{k}\right),\\
    C_{k}|_{t = 0} = 0, & \quad C_{k}|_{r = 1} = -\delta_{0k} ,\quad
    \frac{dC_{k}}{dr}\Big|_{r\rightarrow\infty}  = 0.   
    \end{split}
\end{align}
Similarly, substitute~\eqref{eq:gradientexpand} into~\eqref{eq:ade_gradient} to get
\begin{align}
\begin{split} \label{eq:spectral_c_k_gradient}
    \frac{\partial \tilde{C}_k}{\partial t} + \sum_{n=1}^{\infty}\sum_{m=0}^{\infty}B_{n} \Bigg[f_{nr}&\bigg(E_{mnk}\frac{\partial \tilde{C}_{m}}{\partial r}+E_{1nk}\bigg) - f_{n\theta} \bigg(F_{mnk}\frac{\tilde{C}_{m}}{r}+F_{1nk}\bigg)\Bigg] \\
    &= \frac{1}{\rm Pe}\left(\frac{\partial^{2}\tilde{C}_{k}}{\partial r^{2}}+ \frac{2}{r}\frac{\partial \tilde{C}_{k}}{\partial r} - \frac{k(k+1)}{r^2}\tilde{C}_{k}\right),\\
    \tilde{C}_{k}|_{t = 0} = 0,& \quad
    \tilde{C}_{k}|_{r = 1} =  -\delta_{1k}, \quad
    \frac{d\tilde{C}_{k}}{dr}\Big|_{r\rightarrow\infty} = 0.   
    \end{split}
\end{align}
The terms $f_{nr}$ and $f_{n\theta}$ related to the flow field are given by
\begin{equation}
\begin{split}
    f_{nr} = \left(\frac{1}{r^{n+2}}-\frac{1}{r^{n}}\right),\quad
    f_{n\theta} = \left(\frac{n}{r^{n+2}}-\frac{n-2}{r^{n}}\right).
    \label{fnrtheta}
\end{split}
\end{equation}
The coefficients $E_{mnk}$ and $F_{mnk}$ are given by
\begin{equation}
\begin{split}
    E_{mnk} = \frac{2k+1}{2}\int_{-1}^{1}P_{n}P_{m}P_{k}d\mu,\quad
    F_{mnk} = \frac{(2k+1)}{2n(n+1)}\int_{-1}^{1}(1-\mu^{2})P'_{n}P'_{m}P_{k}d\mu.
    \label{EFmnk}
\end{split}
\end{equation}
Equations~\eqref{eq:spectral_c_k_uniform} and~\eqref{eq:spectral_c_k_gradient} are then discretized in space and time. Following \cite{michelin2011, liu2024optimal}, we use a central difference discretization in space and backward Euler in time. We compute the time evolution of the concentration field and associated nutrient uptake. 
The latter, using properties of the Legendre polynomials, can be expressed as 
\begin{align}
    J(t) = J_h(t) + \epsilon \tilde{J}(t) = \left(\frac{\partial C_0}{\partial r}+ \epsilon\frac{\partial\tilde{C}_0}{\partial r} \right)_{r=1}.
\end{align}


\end{document}